\documentclass[%
 reprint,https://www.overleaf.com/project/612d0589533c89fbc9a0ecea
 amsmath,amssymb,
 aps,
]{revtex4-2}

\usepackage{graphicx}% Include figure files
\usepackage{dcolumn}% Align table columns on decimal point
\usepackage{bm}% bold math
\begin{document}

\preprint{APS/123-QED}

%\title{Particle tracking for correlated motion}% Force line breaks with \\
\title{Improved tracking of particles with highly correlated motion}

\author{Ella M. King$^1$, Zizhao Wang$^2$, David A. Weitz$^{1, 2}$, Frans Spaepen$^2$, Michael P. Brenner$^{1, 2}$}

\affiliation{$^1$ Physics Department, Harvard University}
\affiliation{$^2$ School of Engineering and Applied Sciences, Harvard University}
% \author{Ella King}
%  %\altaffiliation[Also at ]{Physics Department, XYZ University.}%Lines break automatically or can be forced with \\
%  \affiliation{Physics Department, Harvard University}
% \author{Zizhao Wang}%
%  %\email{Second.Author@institution.edu}
% \affiliation{%
%  School of Engineering and Applied Sciences, Harvard University
% }%

% \author{Michael Brenner}
% \affiliation{School of Engineering and Applied Sciences, Harvard University}
% \altaffiliation[Also at ]{
%  Google Brain
% }

\date{\today}% It is always \today, today,

\begin{abstract}
Despite significant advances in particle imaging technologies over the past two decades, few advances have been made in particle tracking, i.e. linking individual particle positions across time series data. The state-of-the-art tracking algorithm is highly effective for systems in which the particles behave mostly independently. However, these algorithms become inaccurate when particle motion is highly correlated, such as in dense or strongly interacting systems. Accurate particle tracking is essential in the study of the physics of dense colloids, such as the study of dislocation formation, nucleation, and shear transformations. Here, we present a new method for particle tracking that incorporates information about the correlated motion of the particles. We demonstrate significant improvement over the state-of-the-art tracking algorithm in simulated data on highly correlated systems.

\end{abstract}

\maketitle

%\tableofcontents

\section{\label{sec:introduction} Introduction}
 Accurately tracking particle positions over time remains a major challenge in advancing colloidal physics research. Single-particle level tracking is critical to studies of the mechanisms of crystal nucleation\cite{rossi2011cubic}, melting\cite{wang2012imaging}, glass formation and dislocations \cite{schall2007structural}, and a variety of other phenomena in dense colloid systems. While advanced confocal microscopy and image analysis techniques have improved the accuracy of the extraction of particle positions from experiments, less progress has been made in accurate tracking of particles exhibiting highly correlated particle motion, such as dense or strongly interacting systems.

 Existing tracking algorithms are based on the fundamental work of Crocker and Grier~\cite{crocker1996methods}, in which positions of particles undergoing Brownian motion are linked between time steps by finding the closest candidates in successive frames. Importantly, this method has two conditions for accurate reconstruction of the tracks. First, the particle displacement between two successive time steps should be much smaller than the inter-particle separation. Second, each particle should diffuse independently. If these two conditions are not met, the trajectory of each particle can often not be distinguished from that of its neighbors. These conditions are violated in systems of strongly interacting particles, which limits the accuracy of the Crocker-Grier algorithm.
 
 Though enhancements to Crocker and Grier's original algorithm have been made, such as handling uniform flow~\cite{kilfoil, allan_daniel_b_2021_4682814}, the fundamental requirement that particle displacement be far less than the inter-particle separation remains a significant limitation. 
 This problem is particularly acute in 3D imaging due to limitations in the frame rate.
 %If the imaging height exceeds several particle diameters, accurate tracking rapidly becomes untenable in systems exhibiting correlated motion. 
 
 In this work, we introduce a new  algorithm that incorporates information about correlations in particle motion, by using simulations to improve estimates of tracking probability distributions. The algorithm achieves up to 33\% more accurate tracking than Crocker and Grier in simulated systems that go beyond the bounds of the algorithm introduced by Crocker and Grier.
 We also demonstrate robust results on noisy experimental data.

\section{Tracking Correlated Motion \label{sec:algorithm}}

\subsection{\label{sec:cg} The Crocker-Grier Algorithm}
% The Crocker-Grier algorithm is a very widely used method for tracking experimental particle data. 
The core assumption of the Crocker-Grier tracking algorithm is that a given particle at position $x_0$ will be found at a position $x$ after a time $t$ has passed with probability \begin{equation}
    P(x, t) = \frac{1}{(4\pi Dt)^{\frac{d}{2}}}e^{-\frac{(x-x_0)^2}{4Dt}}
\end{equation}
where $D$ is the diffusion coefficient and $d$ is the dimension. This is the Green's function solution to the single-particle diffusion equation~\cite{crocker1996methods}. The tracking algorithm is accurate for systems in which particles diffuse independently. However, for systems in which the motion in a group of particles is correlated, the assumption underlying Crocker-Grier breaks down. 

In particular, once multi-particle correlations are introduced, the single-particle diffusion equation is no longer an accurate model of the system. 
% As a result, a new probability distribution is needed to describe the motion of the particles. We therefore introduce a new tracking algorithm that uses information about correlations in the system to improve tracking accuracy.

\subsection{\label{sec:corr} The Correlation Tracking Algorithm}
This new tracking algorithm approximates multi-particle correlations using simulated data to inform predictions. We refer to this method as ``Correlation Tracking.'' Inspired by prior work showing that particle interactions can directly tune dynamics \cite{goodrichking}, we use prior knowledge about particle interactions to make more accurate predictions about the dynamics. By incorporating this physical information into the algorithm, we are able to increase tracking accuracy.

Consider first a pair of frames, $x_0$ and $x_1$. The frame $x_i$ contains a list of particle positions at time $i$. The aim of particle tracking is to match particle labels between frame $x_0$ and frame $x_1$. Because we know the interaction potential, we can simulate the motion beginning at position $x_0$ and let it evolve for the same length of time as between the pair of frames. These simulations are stochastic, and thus we run them multiple times to extract $M$ replicates that approximate a distribution of potential particle positions at time $x_1$. Using this set of potential particle positions, we fit multivariate Gaussian distributions to subsets of the particles. 

Consider a set of $k$ particles (the $k-1$ particles nearest to a central particle) in frame $x_0$. These particles are in a bath of $N-k$ particles.  To track these $k$ particles in the subsequent frame, we perform the simulations described above. We refer to the resulting simulations as $x_{1, \text{sim}}$. All $N$ particles are simulated in each of the $M$ replicates. The $M$ replicates provide mean positions $\mu_i = \langle x_{1, \text{sim}}^{i}\rangle$ for each particle $i$ in the set of $k$ particles. Additionally, we compute correlations between each particle in the set of $k$: $$C_{ij} = \langle (x_{1, \text{sim}}^{i} - \mu_i) \cdot (x_{1, \text{sim}}^{j} - \mu_j) \rangle$$ Together, the computed means and correlations populate a multivariate Gaussian distribution that describes the probability of finding a set of $k$ particles:
\begin{equation}
    P(x | x_0, \alpha) = \frac{1}{\sqrt{(2\pi)^d |C|)}} \exp{-\frac{1}{2}(x - \mu)^TC^{-1}(x - \mu)}
\end{equation}
\label{morse}
where $x$ is the set of positions of the $k$ particles we are interested in tracking, $\alpha$ is the set of parameters describing the interaction potential, $d$ is the dimension, and $\mu$ and $C$ are defined above.

The probability function $P(x | x_0, \alpha)$ differs for each set of $k$ particles. To link particles in one frame to the next, we compute the probability function for each set of $k$ particles and use it to determine which set of $k$ particles in the next frame is most likely to be the same set. Because it is computationally intractable to test every possible ordering of $k$ particles in the frame, we select a set of candidate particles to test. These candidate particles are the $p$ closest particles to the set of $k$ in the first frame.

\section{Results on simulated data \label{sec:sim_results}}

We demonstrate our method in a several simulated systems. We focus on systems that exhibit long-range attraction, as these are systems that do not meet the conditions for accurate tracking with the Crocker-Grier algorithm.

\begin{figure}[h!tpb]
    \centering
    \includegraphics[width=\linewidth]{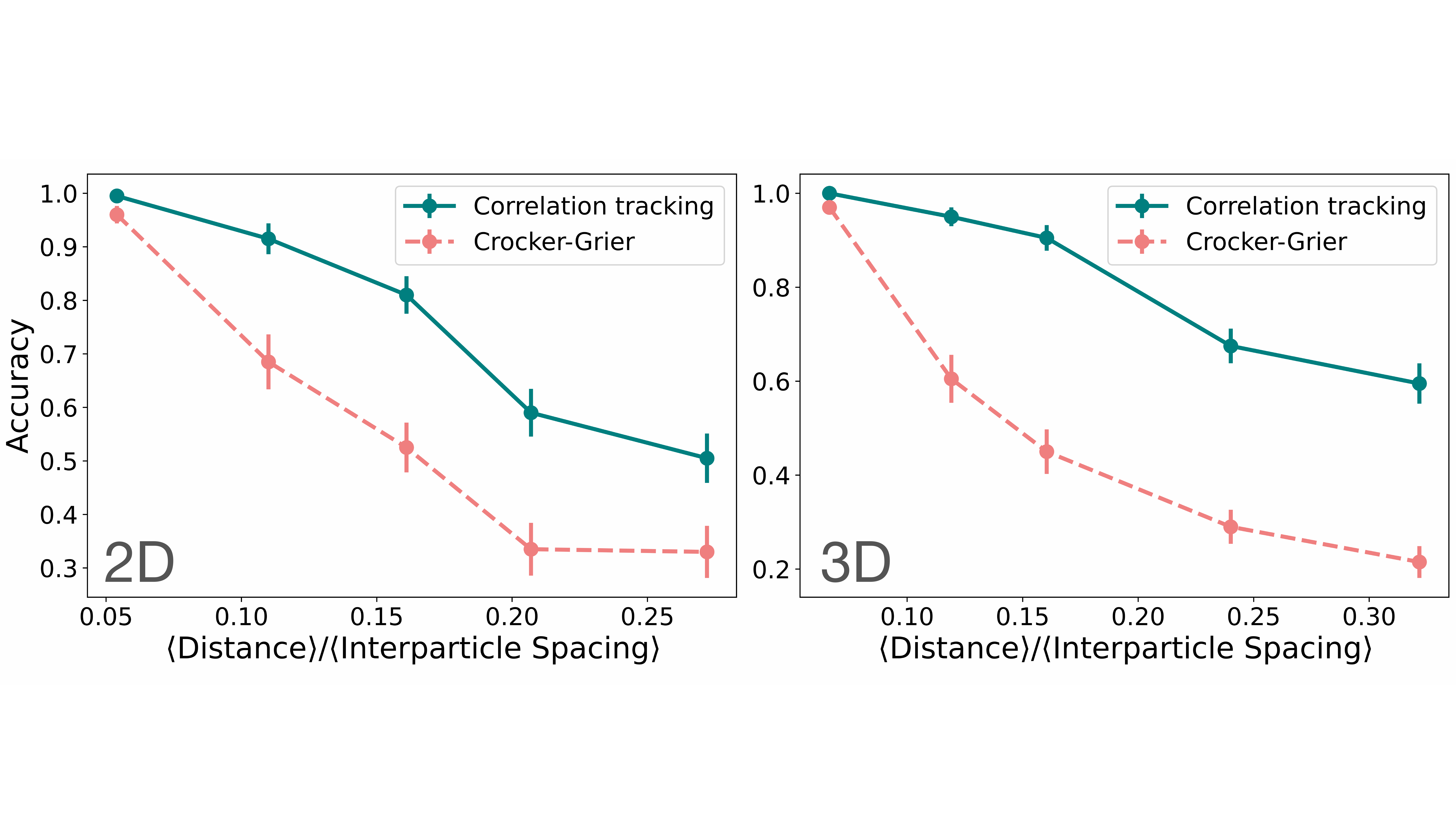}
    \caption{Comparison of the Correlation Tracking results to those of the Crocker-Grier algorithm in simulated data using a Morse pair potential (eq.~\ref{eq:morse}).  The x-axis shows the mean particle displacement divided by the mean inter-particle separation in units of $r_0$. This metric serves as a proxy for the time and enables ready comparison between simulation and experiment. The y-axis shows the tracking accuracy, or the fraction of correctly identified particle labels. The uncertainties are standard errors, based on 25 independent simulations.
    }
    \label{fig:2d_vs_3d}
\end{figure}

In Fig.~\ref{fig:2d_vs_3d}, we show the results of the Correlation Tracking method when applied to a Morse pair potential of the form: 
\begin{equation}{\label{eq:morse}}
    U(r) = D(e^{-2\alpha (r - r_0)} - 2e^{-\alpha(r-r_0)})
\end{equation}
with $D=1$, $r_0=1$ and $\alpha=1$. In all the data shown below, the algorithm was tested on simulated data with 100 particles undergoing Brownian motion with periodic boundary conditions. The volume (or area) fraction of the system was set to 0.5, with $r_0$ as the particle diameter. Molecular dynamics simulations were performed using Jax-MD. When tracking sets of $k$ particles with $p$ candidate particles, we specified $k=5$ and $p=15$ for all data. To estimate the correlation matrix for a given pair of frames, we averaged over 10,000 independent simulations.

As shown in Fig.~\ref{fig:2d_vs_3d}, we observe significant improvement in the tracking accuracy for longer gaps between frames over the state-of-the-art algorithm in both two and three dimensions.
%In three dimensional experimental systems, the $z$-scan creates large time gaps between successive frames. Our method demonstrates considerable increases in accuracy for longer gaps between frames.

\begin{figure}[h!tpb]
    \centering
    \includegraphics[width=\linewidth]{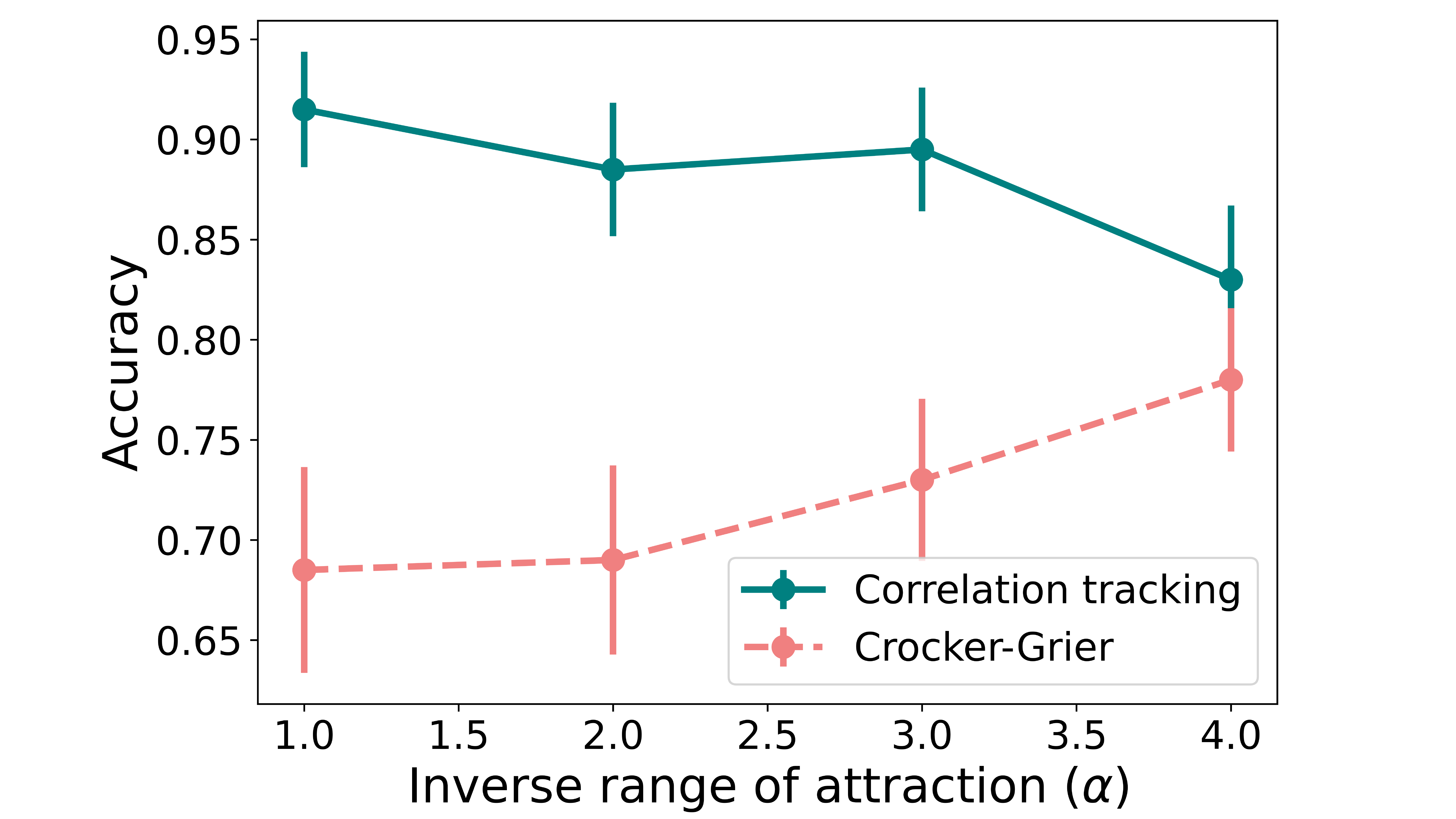}
    \caption{Comparison of the Correlation Tracking algorithm to the Crocker-Grier algorithm as a function of interaction range. The results are computed on simulated data using a Morse potential (eq.~\ref{eq:morse}). The x-axis shows the parameter $\alpha$ in the Morse potential which varies inversely with interaction range. The y-axis shows the fraction of correctly identified particle labels. Error bars are based on 25 independent simulations.  %We demonstrate that a larger interaction range leads to better performance by the Correlation Tracking method and worse performance by the Crocker-Grier algorithm, as this extends beyond the physical bounds in which the Crocker-Grier algorithm is expected to be accurate. When the range of interaction is sufficiently small, the accuracies of the Correlation Tracking algorithm and the Crocker-Grier algorithm converge. 
    }
    \label{fig:range_of_attr}
\end{figure}

We hypothesize that the success of our algorithm in this system is owed to the large range of interaction. The greater the interaction range, the more correlations can be established. To demonstrate this, we compare in Fig~\ref{fig:range_of_attr} the accuracy of our method to that of the Crocker-Grier algorithm as a function of the range of interaction, which is varied by tuning the parameter $\alpha$ in the Morse potential of eq.~\ref{eq:morse}. Note that $\alpha$ varies inversely with interaction range. The figure shows that a larger interaction range indeed leads to improved performance by the Correlation Tracking method and worse performance by the Crocker-Grier algorithm. When the range of interaction is sufficiently small, the accuracies of the Correlation Tracking algorithm and the Crocker-Grier algorithm converge. 
%The Correlation Tracking algorithm matches the accuracy of the Crocker-Grier algorithm for short-ranged attraction and exhibits considerable improvement at longer ranges.

\section{Results on Experimental Data\label{sec:experimental}}
To verify the behavior on experimental data, we tested our tracking algorithm on experimental hard sphere colloids. In this system,  since there are no long-range interactions and the motion is largely uncorrelated,
%diffusive {\bf Not sure what you mean by diffusive}, 
we expect that the algorithm should match the performance of the Crocker-Grier algorithm. 

The experimental system consists of core-shell colloidal particles undergoing Brownian motion in a density-matched and refractive-index-matched solvent. The core and shell of the particles have the same composition: poly trifluoroethyl methacrylate (pTFEMA) and poly tert-butyl methacrylate (pTBMA). The cores are fluorescently dyed. The core diameter is 0.8 $\mu$m, and the particle diameter is 1.75 $\mu$m. This core-shell design separates fluorescent spots in confocal imaging and ensures accurate particle locating in a dense colloidal suspension. A polyelectrolyte brush is grown on the particle surface to prevent aggregation. The particles are suspended in a mixture of urea and formamide, with tetrabutylammonium bromide (TBAB) added as salt to screen charges and to ensure that the particle interaction closely approximates that of hard spheres. This colloidal solution is sealed in a glass chamber with dimensions 20 mm x 10 mm x 0.4 mm. The particles take up 43\% of the total volume.    

\begin{figure}[h!tpb]
   \centering
    \includegraphics[width=0.8\linewidth]{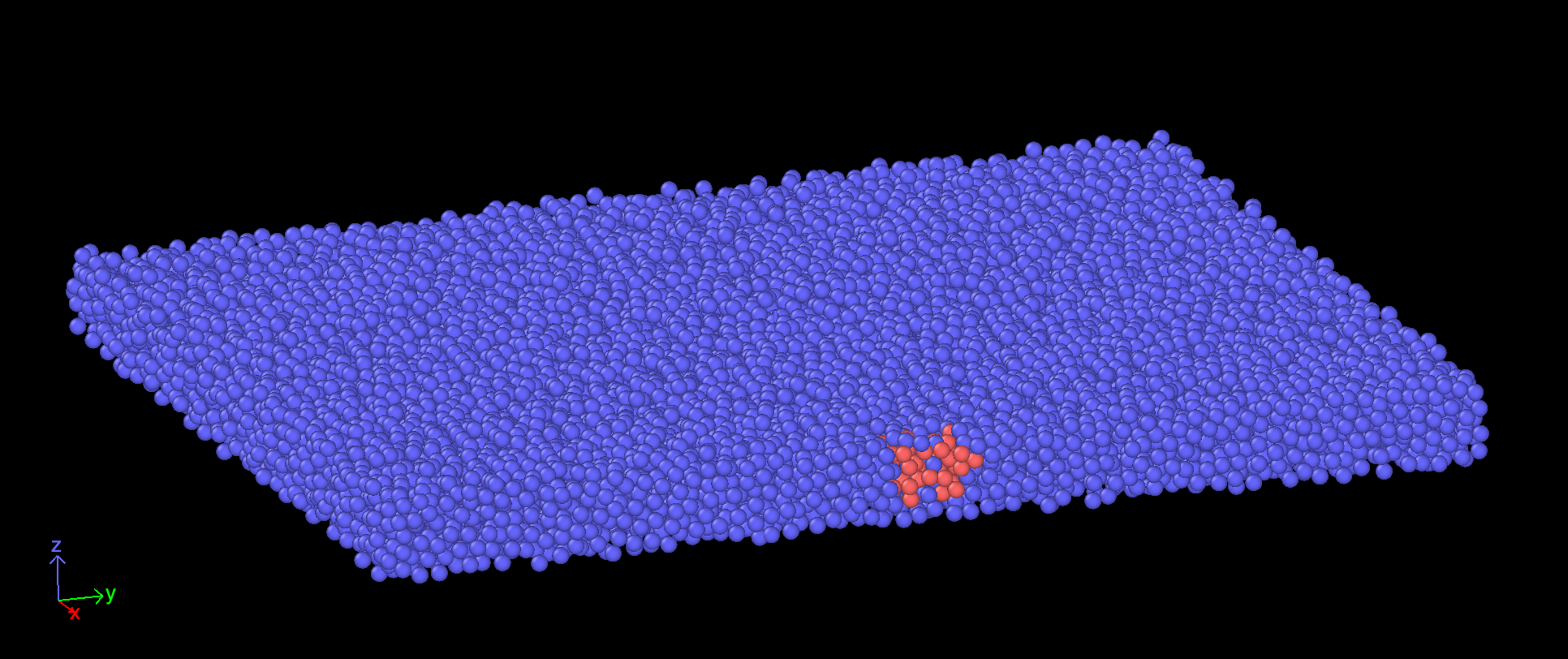}
    \caption{3D rendered image of the colloidal particles in the first frame of the tracking experiment. The front half of the sample has been removed to reveal the red particles at the center, which are used for the tracking analysis. The particle diameter is 1.75 $\mu$m.
    }
    \label{fig:exp_setup}
\end{figure}

We use fluorescent confocal microscopy to image the particles.  Fig.~\ref{fig:exp_setup} shows part of the imaged volume, which measures 209 $\mu$m x 118 $\mu$m x 10 $\mu$m and contains about 38000 particles. The imaged volume is far from the sample boundaries to prevent edge effects. The particle centers are located and tracked using TrackPy\cite{allan_daniel_b_2021_4682814}, a python package based on the Crocker-Grier algorithm. We chose 55 particles at the center of the image for analysis, marked in red in Fig.~\ref{fig:exp_setup}. This ensures that the analyzed particles are present through all 100 frames. The frame rate is 1.25 seconds per 3D stack. This high frame rate allows accurate tracking by the Crocker-Grier algorithm. We compare frames at different time intervals to test changes in tracking accuracy.

\begin{figure}[h!tpb]
   \centering
    \includegraphics[width=0.8\linewidth]{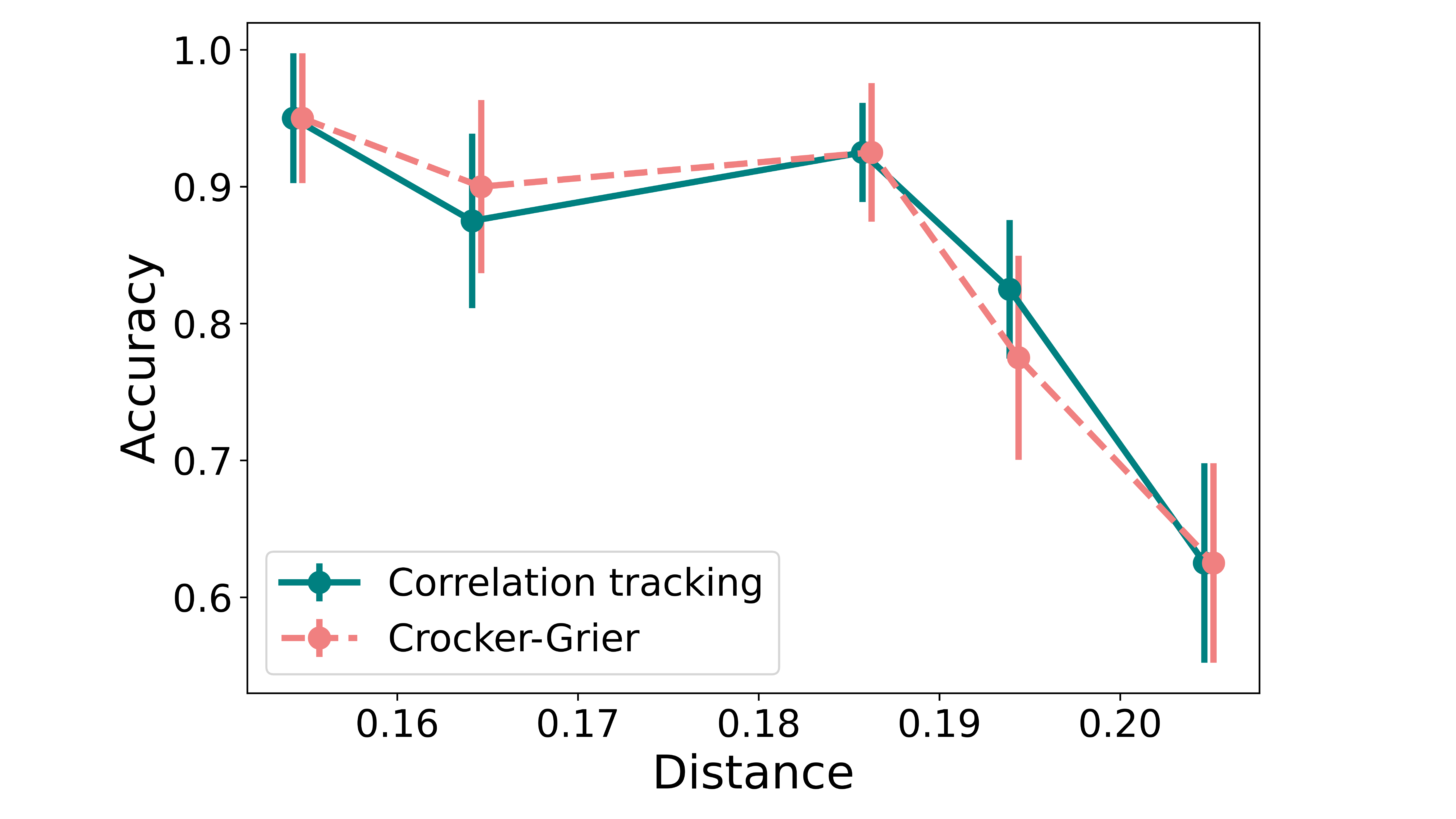}
    \caption{Comparison of Correlation Tracking to Crocker-Grier in a three-dimensional experimental hard sphere system. The x-axis shows the mean particle displacement in units of the particle diameter.
    %per average inter-particle separation. 
    The y-axis shows the fraction of correctly labeled particles. The error bars show the standard error, computed for 10 independent tracking events that use different central particles and different starting frames. 
    }
    \label{fig:exp}
\end{figure}

We approximate the hard sphere behavior in the simulations with a WCA potential \cite{wca}, which consists of the repulsive portion of a Lennard-Jones potential. We use this potential instead of an exact hard sphere interaction to simplify the calculation of the dynamics. We first simulate an ensemble of trajectories starting from the positions in the first frame of the experimental data. This ensemble is used to parameterize a probability distribution over particle positions at a later time. We then label the particles at this later time using the probability distribution we have computed from the simulations.
%Despite an imperfect description of the potential, we are indeed able to recover Crocker-Grier's performance with our algorithm in this regime. Moreover, we perform no noise modeling to move from simulated to experimental data, suggesting our results are robust. 
Fig~\ref{fig:exp} shows the results.

\section{Results on colored particles {\label{sec:colors}} }
\begin{figure}[h!tpb]
   \centering
    \includegraphics[width=\linewidth]{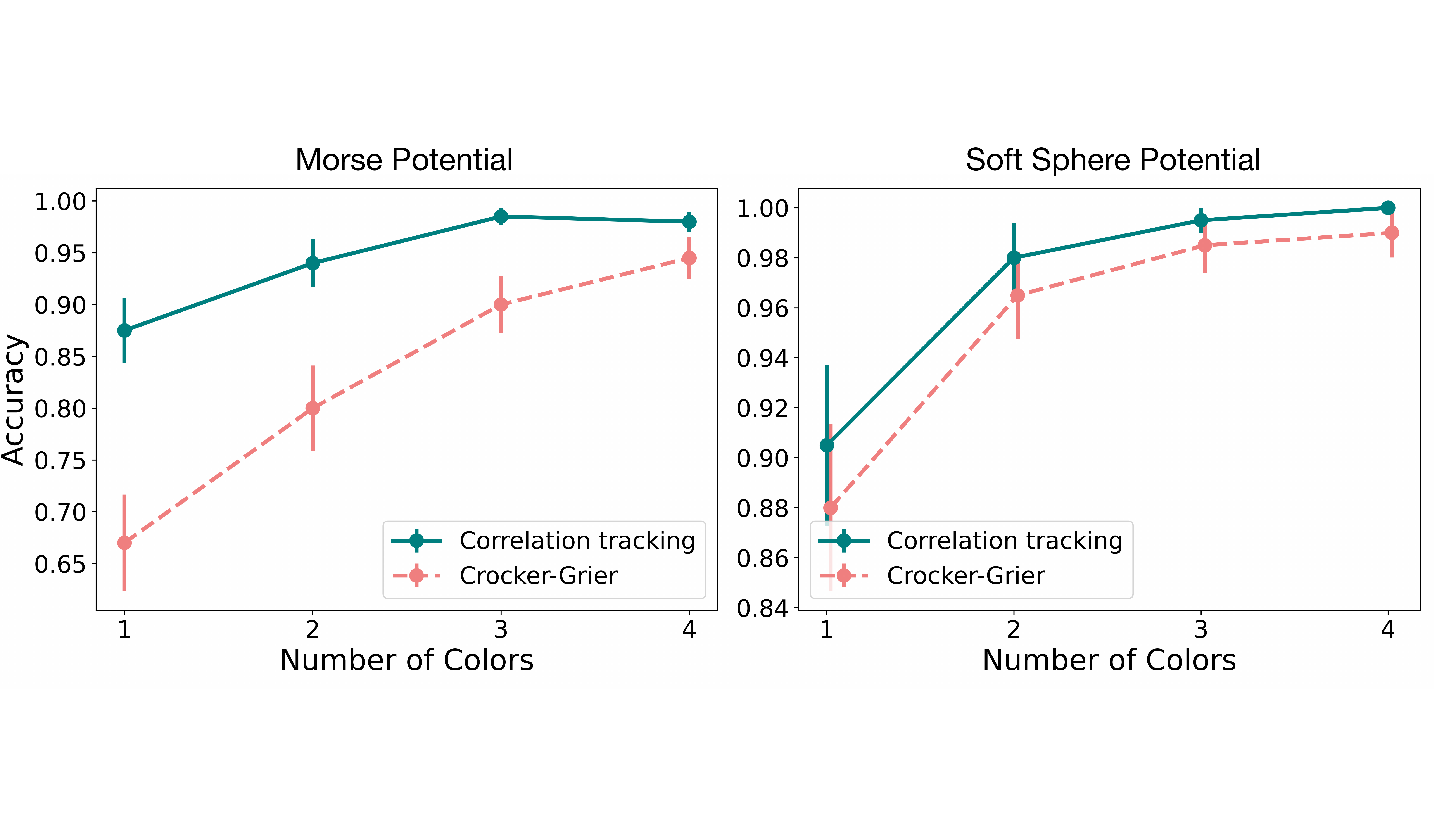}
    \caption{Comparison of the Correlation Tracking algorithm to the Crocker-Grier algorithm in simulated data on particles with multiple colors, interacting via a Morse potential (left) and a soft sphere potential (right). The particles were allowed to evolve such that their distance, measured in units of $r_0$ per average inter-particle separation, was 0.11 for the Morse potential, and 0.069 for the soft sphere potential. The y-axis shows the fraction of correctly labeled particles in that frame. Error bars are based on 25 independent simulations.
    }
    \label{fig:multicolor}
\end{figure}

Lastly, we consider systems with multiple particle types, such as colloids of different colors. Knowledge of these types can be used to inform the tracking algorithm and offer more accurate predictions. First, consider the case of $N$ particles interacting by a Morse potential. Suppose each particle is dyed one of $M$ possible colors. If $M=1$, we recover the results found above. As $M$ approaches $N$, all the particles become distinguishable, and it is always possible to track every particle accurately. In the intermediate range, more colors lead to an effectively reduced density without interfering with correlated particle motion. 
% Though the Crocker-Grier algorithm is expected to perform well in low-density systems, it is not expected to be accurate for highly correlated motion. On the other hand, the Correlated Tracking algorithm should perform well in cases of highly correlated motion.
Thus, noting that introducing $N$ independent colors is rarely experimentally feasible, we examine the tradeoff between limiting the number of colors and the accuracy of tracking using both the Correlated Tracking algorithm and the Crocker-Grier algorithm in Fig.~\ref{fig:multicolor}.

To simulate adding independent colors, we added the concept of species  to our simulation. We assume that if a particle is dyed a given color, it may only be mapped to a particle of the same color in the next frame. To enforce this constraint, we treat each color  as a different species and track it separately, which prevents errors in labeling across colors in both the Correlation and Crocker-Grier tracking algorithms. In this example, the species argument is only relevant to the tracking analysis, and not to the simulations themselves; here we assume that differently colored particles interact by a Morse potential in the same way as particles with the same color.
%{\bf Reworded--of course different species could interact differently}

Fig.~\ref{fig:multicolor} shows that increasing the number of colors increases tracking accuracy for both algorithms. Adding too many colors leads to diminishing returns in tracking accuracy. For the Morse potential, Crocker-Grier achieves very high accuracies with 4 colors.  Correlation Tracking improves this and achieves nearly perfect tracking accuracy with only 3 colors. For the soft sphere potential, Crocker-Grier achieves high accuracy with 3 colors. The Correlation Tracking algorithm matches the same level of accuracy Crocker-Grier achieves with only 2 colors.

\section{Discussion \label{sec:discussion}}

Our results on experimental data indicate that the Correlation Tracking algorithm is robust. Even though we used an imperfect description of the hard sphere potential in the experimental system, the algorithm  recovers Crocker-Grier's performance in the diffusive regime. Many algorithms, particularly machine learning algorithms, require extensive modeling of noise sources in the experimental system to achieve accurate results on experimental data \cite{yue2019variational, 8363578, gupta2019dealing}. For instance, computer vision algorithms often require modeling noise due to the type of camera and the method of transmitting data \cite{yue2019variational}.
%{\bf What do you mean by noise modeling? Which noise are they modeling?}
In contrast, we are able to accurately track experimental data without introducing any noise modeling. Therefore, our model is robust to noise sources in the experimental data, such as polydispersity and positional uncertainty. Furthermore, the Correlation Tracking method relies on an accurate description of the potential. However, we noted that our description of the potential is imperfect, yet we are still able to track particles accurately. As long as the potential is accurate over the range and timescale of the measurements, Correlation Tracking will be accurate.
%on the timescale of the frame rate, Correlation Tracking will be accurate.
%{\bf I still worry about whether the potentials themselves are accurate. This isn't noise -- it is uncertainty. Can we say anything about how accurate the potentials have to be for the method to work?  This is the weakest part of the method--the need to accurately know interactions..}

The increase in tracking accuracy due to the Correlation Tracking algorithm comes at a significant computational cost. We mitigate that cost by taking advantage of the automatic vectorization and compilation offered by the package JAX \cite{jax2018github, jaxmd2020}, so that simulations are run simultaneously. Tracking a set of 5 particles across one pair of frames with a batch size of 10,000 simulations used to estimate the correlation matrix takes approximately 4 minutes on a K80 GPU on Google Colab, whereas performing the same analysis with the Crocker-Grier method takes only a few seconds. However, the proof of principle of the Correlation Tracking algorithm leaves a lot of room for improvements in computational efficiency. Moreover, for commonly used implementations of the Crocker-Grier algorithm, such as TrackPy, the computational cost grows factorially when the number of particles in a tracking set increases. This makes tracking in a dense particle system particularly costly in computation, because many candidate particles need to be considered in a tracking set.

%{\bf Should this paragraph move up to your discussion of computation time?} 
Though our method increases computation time, it may reduce the overall experimental effort. As discussed in section \ref{sec:sim_results}, one method of increasing tracking accuracy is to introduce multiple colors for particles in an experimental system. Creating each color, however, comes at a high cost in experiment time. Since the Correlation Tracking method significantly reduces the number of colors needed to reach the highest levels of tracking accuracy, it may save significant time.
%% (original) Moreover, commonly used implementations of the Crocker-Grier algorithm, such as TrackPy, offer other features that also increase its computational cost considerably.

The Correlation Tracking algorithm uses simulation in a fundamentally new way. To date, simulations have been used to discover features and mechanisms of material and biological systems that cannot yet be observed in experiment\cite{porter2021shear, du2017shape}, to deepen our understanding of experimental results \cite{li2021microscopic, audus2021molecular}, or to guide future directions for experimental exploration \cite{kimchi2020self}. This algorithm, however, intertwines simulation and experiment in the data analysis procedure. The simulated data is used to directly increase the accuracy of particle tracking by rooting the data analysis method in the physics of the system. There is a rich space to explore by bringing simulation and experiment in direct contact, a space that we anticipate will contain new lines of research on both sides.

As the Correlation Tracking algorithm increases tracking accuracy, it allows accurate tracking for longer image acquisition intervals in experiments. This enables a taller imaging region in 3D confocal imaging. For example, in the experiments of this paper, the microscope can reach 40 frames per second while maintaining the necessary image resolution in a xy-plane for accurate particle locating. The step size in the z-direction needs to be 0.18 $\mu$m for accurate particle locating in the z dimension. The dynamics of the colloidal system limit the 3D imaging interval to less than 1.5 seconds to ensure accurate particle tracking using the Crocker-Grier algorithm. These restrictions give a maximum of 60 xy-planes or 10.8 $\mu$m at each time step. With this limitation, only 7 particle layers can be scanned for the 1.75-$\mu$m-sized particles used in this experiment. A significant portion of the particles moves out of the imaging region in a longer experiment. This makes it difficult to track each particle over a long time and to study kinetic phenomena. However, with Correlation Tracking, more particle layers can be imaged at each time step, and long-time particle tracking becomes more feasible. 
%{\bf Is this something we could demonstrate explicitly with the experimental data -- it is a bit of an odd conclusion given that the performance of the two algorithms is similar in the figure}

The Correlation Tracking method promises to be of greatest use in studies with highly correlated motion, such as studies of crystal nucleation or shear thickening in attractive colloidal systems. The algorithm can be applied to any experimental system for which the particle interactions are well-characterized. Moreover, recent work has had significant success in approximating interaction potentials from simulated data using machine learning algorithms \cite{vandermause2021active, chan2019machine}. If these interaction potentials were learned from experimental data, the Correlation Tracking algorithm could be used to provide accurate tracking. These approximations need only be accurate over short distance and time scales to be useful for the Correlation Tracking algorithm. Further, the Correlation Tracking algorithm could be inverted to provide estimates of interaction potentials in experimental systems: given data with accurate particle tracks, the Correlation Tracking algorithm would provide a probability that the tracks are well explained by a given interaction potential. One could converge on an accurate potential by proposing a given interaction potential, evaluating the probability that it explains the experimental particle tracks, and accordingly updating the proposed potentials, for example with gradient-based optimization methods.

\section*{Acknowledgements}
We thank Peter Lu for many valuable discussions and experimental guidance. This material is based on work supported by the National Science Foundation Graduate Research Fellowship under Grant No. DGE1745303,  the Harvard Materials Research Science and Engineering Center (DMR 20-11754),  the Office of Naval Research (ONR N00014-17-1-3029) and the Simons Foundation.

%\nocite{*}

\bibliography{biblio}% Produces the bibliography via BibTeX.

%\appendix

%\section{Experimental Methods \label{sec:exp_supp}}
%Put experimental details here

\end{document}